\documentstyle[11pt]{article}
\normalbaselines
\begin{document}
\bibliographystyle{unsrt}
\def\beq{\begin{equation}}
\def\eeq{\end{equation}}
\vbox{\vspace{38mm}}
\begin{center}
{\LARGE \bf GEOMETRICAL PHASE,GENERALIZED QUASIENERGY AND FLOQUET OPERATOR
AS INVARIANTS
}\\[5mm]

V.I.Manko\footnote{On leave from Lebedev Institute of Physics, Moscow,
Russia.}
\\{\it Dipartimento di Scienze Fisiche, Universita di Napoli,
Italy.}\\[5mm]
 \end{center}

\begin{abstract}
For time-periodical quantum systems generalized Floquet operator is
found to be integral of motion.Spectrum of this invariant is shown to be
quasienergy spectrum.Analogs of invariant Floquet operators are found for
nonperiodical systems with several characteristic times.Generalized quasienergy
 states are introduced for these systems.
Geometrical phase is
shown to be integral of motion.

\end{abstract}

Quantum mechanics is usually connected with energy spectra of the systems
with stationary
Hamiltonians.The dynamics of these systems is described by the transitions
between the energy levels.The nonstationary quantum systems
have no energy levels due to the absence of symmetry related to time
displacements.But for periodical quantum systems
 there exists the symmetry corresponding to
crystal time structure of Hamiltonian.Due to this the notion of quasienergy
levels has been introduced in Ref.\cite{zel67} and \cite{rit67}.The main point
of the quasienergy concept is to relate the quasienrgies to the eigenvalues of
the Floquet operator which is equal to the evolution operator of a quantum
system taken
at a given time moment.The aim of this article is to relate the Floquet
operator to integrals of motion and to introduce a new operator which is
the integral of motion and has the same quasienergy spectrum that the Floquet
operator has.Implicitly this result was contained in Ref.\cite{mal79}
but we want to have the explicit formulae for the new integral of motion.\\
If one has the system with hermitian Hamiltonian $H(t)$ such that
$H(t+T)=H(t)$ the unitary evolution operator $U(t)$ is defined as follows
\beq
|\psi,t>=U(t)|\psi,0>
\eeq
where $|\psi,0>$ is a state vector of the system taken at the initial
time moment.Then by definition the operator $U(T)$ is called the Floquet
operator and its eigenvalues have the form
\beq
f=exp(-iET)
\eeq
where $E$ is called the quasienergy and the corresponding eigenvector is
called the quasienergy state vector.The spectra of quasienergy may be either
discrete or continuous ones (or mixed) for different quantum systems.
For multidimensional
systems with quadratic Hamiltonians the quasienergy spectra have been related
to real symplectic group $ISp(2N,R)$ and found
explicitly in Ref.\cite{mal79}.We want
to answer the following questions.Is the quasienergy conserved observable?
This question is related to another one.Is the Floquet operator $U(T)$ the
integral of motion?The operator $U(T)$ does not satisfy the relation
\beq
dI(t)/dt+i[H(t),I(t)]=0,(\hbar=1)
\eeq
which defines the integral of motion $I(t)$.So,the Floquet operator $U(T)$
is not the integral of motion for the periodical nonstationary quantum
systems.But as it was pointed out in Ref.\cite{mal79} any operator of the
form
\beq
I(t)=U(t)I(0)U^{-1}(t)
\eeq
satisfies the equation (3) and this operator is the integral of motion for
the quantum system under study.Let us apply this ansatz to the case of
periodical quantum systems.We introduce the unitary operator $M(t)$ which
has the form
\beq
M(t)=U(t)U(T)U^{-1}(t).
\eeq
This operator is the integral of motion due to the construction given
by the formula (4) for any integral of motion.The spectrum of the new invariant
operator $M(t)$ coincides with the spectrum of the Floquet operator $U(T)$.We
have proved that since quasienergies are defined as eigenvalues of the
integral of motion $M(t)$ they are conserved quantities.\\The given
construction permits us to introduce new invariant labels for nonperiodical
systems,for example, with the time-dependence of the Hamiltonian
corresponding to quasicrystal structure in time described by two (or more)
characteristic times.For such systems the analog of the invariant Floquet
operator (5) will be given by the relation
\beq
M_{1}(t)=U(t)U(t_{1})U(t_{2})U^{-1}(t).
\eeq
For polydimensional structure charactrised by $N$ times $t_{1},t_{2},...,t_{N}$
we can introduce the integral of motion
\beq
M_{2}(t)=U(t)U(t_{1})U(t_{2})...U(t_{N})U^{-1}(t).
\eeq
The eigenvalues of the operators $M_{1}(t)$ and $M_{2}(t)$ are the consrved
quantities and they characteryse the nonperiodical quantum systems with
quasicrystal structure in time in the same manner as quasienergies describe
the states of periodical quantum systems.\\So,for stationary systems the
spectrum of evolution operator $U(t)$ determines the spectrum of the
Hamiltonian and vice versa.For periodical systems the spectrum of the evolution
operator plays more important role.It determines the quasienergies and,in fact,
the geometrical phase of the system which is the characteristic of special
nonperiodical
system for which the parameters of the Hamiltonian take their initial values
after some time $T$ \cite{ber84},\cite{boh89}.
We will see below that the quasienergy for periodical systems and the
geometrical phase for nonstationary quantum system are the identical objects
\cite{sim90}.
For the systems which are characterised by more complicated time structure
the evolution operator taken at characteristic points and the spectra of
the operators of the type (6) and (7) are the invariant characteristics
of the systems generalizing the notion of quasienergy for nonstationary
quantum systems with deformed symmetry properties.To find generalized
quasienergy states for the nonperiodical system with the integral of motion
(7) we have to find the eigenvalues and the eigenstates of this invariant.
The eigenvalue of the invariant (7) has the form (2) but in the case of
nonperiodical systems we will call the number $E$ the generalized
quasienergy and the corresponding eigenvector will be called the
generalized quasienergy state vector.The operator (7) may be represented in
the form
\beq
M_{2}(t)=exp(-iTH_{ef}(t))
\eeq
where $H_{ef}(t)$ is the invariant effective Hamiltonian.Its spectrum
provides the generalized quasienergy levels of the nonperiodical system.
\\The quasienergy spectrum of periodically kicked quantum systems may
be connected with quantum chaos phenomenon (see Ref.\cite{haa91},
\cite{chi79},\cite{cas88}).In the Ref.\cite{haa92} the integral of motion
for delta-kicked nonlinear oscillator has been found to exist even
in the case of chaotic behaviour.In the Ref.\cite{kar91} the symmetry group
criterium for the periodically delta-kicked systems has been found to
obtain either regular or chaotic behaviour of these systems.The criterium
relates the Floquet operator spectrum to the conjugacy class of the system
symmetry group.For the quadratic hermitian Hamiltonians this group turnes
out to be the real symplectic group $Sp(2N,R)$.The results of the
Ref.\cite{kar91} may be applied to the nonperiodical systems too.
The generalized quasienergy spectrum is determined by the conjugacy class
of the same group to which belongs the integral of motion (7).For quadratic
systems it is the same real symplectic group $Sp(2N,R)$.In the case of two
characteristic times the classification of the possible either chaotic or
regular regimes of the system under study coincides formally
with the classification
for the delta-kicked periodical quantum systems given in
Ref.\cite{kar91}.\\If the Hamiltonian of the nonstationary system has the
symmetry property
\beq
H(t+iT)=H(t)
\eeq
where $T$ is a real number the loss-energy states exist as analogs of
quasienergy states for periodical systems (see Ref.\cite{dod78}).The
construction of generalized quasienergy states given in this talk may
be repeated also for the nonperiodical quantum systems with two (or more)
imaginary characteristic times.For systems with quasicrystal structure
in time but with the imaginary characteristic times the generalized
loss-energy states may be constructed.The mixed situation when some of
the characteristic times are real numbers and the others are the imaginary
ones may be of special interest.
\\The geometrical phase
is defined as the phase of eigenvalue of the evolution operator $U(T)$ where
$T$ is the time moment at which the parameters of Hamiltonian take their
initial values.We answered the same question as for the quasienergies.Is
the geometrical phase the integral of motion of the nonstationary and
nonperiodical quantum system?The answer is "yes" because such nonperiodical
system has characteristic time $T$.Due to this the operator $U(t)U(T)U^{-1}(t)$
is the integral of motion of the system.The form of this invariant coincides
with the form of the invariant (5) of periodical systems but the physical
meaning is different.The phase of eigenvalue of this integral of motion
is the integral of motion too.But this phase is the geometrical phase of
the quantum system since the spectrum of the operator $U(T)$ and the
spectrum of the operator $U(t)U(T)U^{-1}(t)$ coincide.Thus we have proved
that the geometrical phases of nonstationary systems are integrals of motion
for these systems.The integrals of motion found in this work are of the same
nature that have the time-dependent integrals of motion for the harmonic
oscillator with varying frequency and a charge moving in varying magnetic
field constructed in Ref.\cite{mal70},\cite{Malkinmankotrifonov69} and
\cite{tri70}.Such integrals of motion have been analyzed and applied to
general problems of quantum mechanics and statistics in Ref.\cite{dod89}.
For periodical system it was shown in Ref.\cite{mal79} that such
time-dependent integrals of motion exist which at the times $nT$
($n=0,1,2...$) take their initial values.The quasienergy states are the
eigenstates of these integrals of motion.The time-dependent invariants
realyze the identity representation of the lattice translation group
in time which is symmetry group of periodical quantum system.The
generalyzed quasienergy states may be also related to the integrals of
motion which commute with invariant Floquet operator for nonperiodical
quantum systems.From that point of view the nonperiodical systems with
several characteristic times have to demonstrate similar physical properties
that are usually considered as properties of purely periodical quantum
systems.Thus,the types of chaotic and regular regimes of the kicked
quantum systems are the same for both periodical and nonperiodiacal
kicks.

\end{document}